\documentclass[fleqn,twoside]{article}
\usepackage{espcrc2}
\usepackage{amssymb}
\usepackage{graphicx}
\usepackage[figuresright]{rotating}

\newcommand{\AmS}{{\protect\the\textfont2
  A\kern-.1667em\lower.5ex\hbox{M}\kern-.125emS}}
\title{Glueballs at finite temperature from AdS/QCD}
\author{Alex S. Miranda\address{Laborat\'orio de Astrof\'{\i}sica
              Te\'orica e Observacional\\ 
              Departamento de Ci\^encias Exatas e Tecnol\'ogicas\\
              Universidade Estadual de Santa Cruz, 45650-000,
               Ilh\'eus, BA, Brazil}\address[UFRJ]{Instituto de
              F\'{i}sica, Universidade Federal do Rio de Janeiro,\\
              Caixa Postal 68528, RJ 21941-972, Brazil}\thanks{email: astmiranda@if.ufrj.br},
        C. A. Ballon Bayona\address{Centro Brasileiro de
              Pesquisas F\'{i}sicas, Rua Dr. Xavier Sigaud 150,\\
              Urca, 22290-180, Rio de Janeiro, RJ, Brazil}\thanks{email: ballon@cbpf.br},
        Henrique Boschi-Filho\addressmark[UFRJ]\thanks{email: boschi@if.ufrj.br}
        and
        Nelson R. F. Braga\addressmark[UFRJ]\thanks{email: braga@if.ufrj.br}}

\begin{document}

\begin{abstract}
Inspired in the AdS/CFT correspondence, a variety of holographic
phenomenological models have been proposed in the last years
to describe non-perturbative aspects of strong interactions. These
models are denominated as AdS/QCD.
In this work we review the use of the AdS/QCD soft-wall model to investigate
the spectrum of scalar glueballs at finite temperature. The scalar glueball states 
are identified as the poles of the retarded correlation function of the glueball operator.
In the gauge/gravity duality, these poles are determined by the quasinormal spectrum of a massless scalar field
propagating in the bulk geometry that consists on an $\mbox{AdS}_{5}$
black hole with a background dilaton field. 
We discuss some results for masses and decay widths of scalar glueballs in the plasma phase 
and analyse how these quantities evolve with temperature and momentum. \vspace{1pc}
\end{abstract}

\maketitle

\section{Introduction}

The strong interactions are described by QCD which is a SU(3) gauge theory.
At high energies, the coupling constant is small and the theory can
be treated with perturbative methods. When one considers processes involving
high energies one finds very good agreement between theoretical
predictions and experimental results. However, many important aspects
of strong interactions, like mass generation and quark confinement,
are not yet completely understood because they involve the low-energy
non-perturbative regime of QCD. So, it is important to look for
alternative tools to deal with this non-perturbative regime. A traditional
approach is to use lattice simulations that still do not furnish a
complete description of low-energy QCD. A more recent proposal is to
construct phenomenological models for strong interactions based on
the AdS/CFT correspondence \cite{Maldacena:1997re,Witten:1998qj,Gubser:1998bc}.
These models are presently known as AdS/QCD. The simplest model,
known as hard wall, consists in using an AdS slice with a size related to an
infrared cut-off in the gauge theory. This model was used to estimate hadron masses \cite{BoschiFilho:2002ta,BoschiFilho:2002vd,deTeramond:2005su,Erlich:2005qh,DaRold:2005zs,BoschiFilho:2005yh}.
It presents confinement at zero temperature \cite{BoschiFilho:2006pe} and deconfinement at high
temperatures \cite{BoschiFilho:2005mw}. Another interesting AdS/QCD model that
leads to linear Regge trajectories is known as soft wall. 
In this model a non-uniform background dilaton field plays the role of an
effective infrared cut-off. This model was also used to calculate hadronic
masses \cite{Karch:2006pv,Colangelo:2007pt,Colangelo:2008us}. The
confinement/deconfinement thermal phase trasition in both models
was studied in Refs. \cite{Herzog:2006ra,Kajantie:2006hv,BallonBayona:2007vp}. 
It was found that there is a Hawking-Page gravitational transition such
that the space with a black hole is thermally favored at high temperatures,
while below some critical (model-dependent) value, the thermally
favored space is a pure AdS.

In this paper we are going to review some recent progress in using the
soft-wall AdS/QCD model to compute the spectrum of scalar
glueballs in the finite-temperature plasma phase of QCD \cite{Miranda:2009uw}
(See also Ref. \cite{Colangelo:2009ra}). We consider the AdS black-hole spacetime 
for all temperature values. As commented above, for temperatures lower
 than a critical value the black hole is thermodynamically unstable and can be interpreted as a
supercooled Yang-Mills plasma. 

The massless scalar field in the 5-d bulk
is dual to scalar glueballs in the 4-d boundary theory and the black-hole 
quasinormal modes correspond to the poles of the retarded correlation function of the operator
${\cal O}=\mbox{Tr}(F^2)$ that creates scalar glueball states. With a numerical analysis,
we obtain the real and imaginary parts of the
quasinormal mode frequencies as a function of the temperature 
and the spatial momentum. The numerical computations at high temperatures
were performed using series solutions of the equation of motion.
At low temperatures, we found problems with the convergence of the series so it was necessary 
to use an alternative approach. We have chosen 
the Breit-Wigner method which has been recently applied to compute
black-hole quasinormal modes in Ref. \cite{Berti:2009wx}.
We calculate the retarded Green's function of scalar glueballs and verify
that their poles coincide with the quasinormal mode frequencies.
The imaginary part of the retarded Green's function gives the spectral
function, which we also analyze in this article.

Other recent results in scalar glueballs can be found in \cite{Forkel:2007ru}. 
For a general review on glueballs see \cite{Mathieu:2008me}.

\section{The soft-wall model at finite temperature}

Our main objective in this paper is to review recent work
\cite{Miranda:2009uw} on the spectrum of scalar glueballs
at finite temperature in the soft-wall AdS/QCD model. For this
purpose, it is considered a dilaton field $\Phi\,=\,cz^2$ in a
finite-temperature background that consists of an AdS black-hole spacetime
with metric 
\begin{equation}
ds^{2}=e^{2A}\left[-f\,dt^2+dx^{i}dx^{i}+f^{-1}\,dz^2\right],
\label{background}
\end{equation}
where $f=1-(z/z_{h})^4$, $A=-\mbox{ln}(z/L)$ and
$L$ denotes the AdS curvature radius. The coordinate $z$ is defined in the range
$0\le z \le z_h$ and $z=0$ corresponds to the boundary of the spacetime. 
The parameter $z_{h}$ indicates the position of the event horizon which 
is related to the black-hole Hawking temperature $T$ by the relation
$z_{h}=1/\pi T$, where $T$ also represents the temperature of the
boundary field theory.

The $\mbox{AdS}_{5}$ black-hole spacetime considered here is
associated to three distinct phases, each one corresponding to a range
of values of the temperature parameter $\widetilde{T}=\pi T/\sqrt{c}$.
The thermal competition between the black hole and thermal AdS space is such
that there is a Hawking-Page transition between spaces at $\widetilde{T}_{c}^{\,2}
\approx 2.38644$. This transition is interpreted in the boundary field theory as a
confinement/deconfinement phase transition. The black hole spacetime
is the dominant configuration for $\widetilde{T}>\widetilde{T}_{c}$,
while the thermal AdS space is the dominant one for low temperature values.
Hence, for $\widetilde{T}<\widetilde{T}_{c}$, the plasma will be in a metastable (unstable)
 phase corresponding to a positive (negative) sign of the black-hole specific heat $\mathcal{C}_{BH}$. 
It was shown in Ref. \cite{Miranda:2009uw} that
$\mathcal{C}_{BH}>0$ for $\widetilde{T}^{\,2}\gtrsim 0.75$
and $\mathcal{C}_{BH}<0$ for $\widetilde{T}^{\,2}\lesssim 0.75\,$.

In the soft-wall model \cite{Karch:2006pv}, the dilaton background
field $\Phi(z)=cz^{2}$  interacts with the bulk fields through
the replacement of the action integrals: $\int d^{5}x\sqrt{-g}\,{\cal L}
\Rightarrow \int d^{5}x\sqrt{-g}e^{-\Phi}{\cal L}$. In particular,
the action for the massless scalar field $\phi$ is given by:
\begin{equation}
S=-\frac{\pi^3 L^5}{4\kappa_{10}^2}\int d^{5}x\sqrt{-g}
e^{-\Phi}\,g^{MN}\partial_{M}\phi\partial_{N}\phi\,,
\label{action0}
\end{equation}
where $g_{MN}$ is the black-hole metric given by (\ref{background}) and
$\kappa_{10}$ is the ten-dimensional gravitational constant.
The indices $M,N$ run over $0,\,1,\,...,4$ where  $x^{\mu}$ ($\mu=0,...,3$)
are the 4$d$ boundary coordinates and $x^{4}=z$ is the extra radial coordinate.

\section{The Schr\"odinger equation}

We do not consider here the backreaction of the dilaton $\Phi$
and of the scalar field $\phi$ on the metric. Then, from the
action (\ref{action0}), we obtain a linearized equation of motion
for $\phi$:
\begin{equation}
\frac{e^{\Phi}}{\sqrt{-g}}\partial_{z}\left(\sqrt{-g}e^{-\Phi}
g^{zz}\partial_{z}\phi\right)+g^{\mu\nu}\partial_{\mu}\partial_{\nu}\phi\,=\,0
\, .\label{eqwave1}
\end{equation}
After Fourier decomposition of $\phi$ with respect to the coordinates 
$x^{\mu}$, Eq. (\ref{eqwave1}) becomes
\begin{equation}
e^{B}f\,\partial_{z}\left(e^{-B}f\,\partial_{z}\phi\right)+
\left(\omega^{2}-f\,q^{2}\right)\phi=0 \, ,
\label{eqwave2}
\end{equation}
where $k_{\mu}=(-\omega,q_{i})$, $\,\,\,q^{2}=\sum_{i=1}^{3}q_{i}^{2}$
and \break $\,B=\Phi-3A=cz^2+3\,\mbox{ln}(z/L)$.

The foregoing equation takes the form of a one-dimensional
Schr\"odinger equation when written in terms of the new
variable $\psi=e^{-B/2}\phi$:
\begin{equation}
\partial_{r_{\ast}}^{2}\psi+\omega^2 \psi=V\psi,
\label{Schrod1}
\end{equation}
where $r_{\ast}$ is the tortoise radial coordinate, defined in 
such a way that $dz/dr_{\ast}=-f$, and
the effective potential $V$ is given by
\begin{eqnarray}
V&=&\frac{f}{z^2}\left[q^{2}z^{2}+\frac{15}{4}+\frac{9}{4}
\frac{z^{4}}{z_{h}^4}\right.\nonumber\\
& &+\,2cz^{2}\left.\left(1+\frac{z^4}{z_h^4}\right)+
c^{2}z^{4}f\right].
\label{potential}
\end{eqnarray}

The general solution of equation (\ref{Schrod1}) can be written as a
linear combination of a normalizable solution $\psi_1$ and a
non-normalizable solution $\psi_2$. These wave functions have the
following asymptotic form near the boundary $z=0$: 
\begin{eqnarray}
\psi_{1}&=&z^{5/2}\left[1+a_{11}z^{2}+a_{12}z^{4}+\cdots\right],
\label{normalizable}\\
\psi_{2}&=&z^{-3/2}\left[1+a_{21}z^{2}+a_{22}z^{4}+\cdots\right]\nonumber\\
&&+\;b\;\psi_{1}\,\mbox{ln}(cz^2),
\label{nonnormalizable}
\end{eqnarray}
where the coefficients above are given by
\begin{eqnarray}
a_{11}&=&-\frac{\omega^2-q^2-2c}{12}\,,\\
a_{12}&=&\frac{(\omega^2-q^2-2c)^2}{12 \times 32}+\frac{1}{2z_{h}^{4}}+
\frac{c^{2}}{32}\,,\\
a_{21}&=&\frac{\omega^2-q^2-2c}{4}\,,\\
b&=&-\frac{(\omega^2-q^2)(\omega^2-q^2-4c)}{32}\,.
\end{eqnarray}
The coefficient $a_{22}$ is arbitrary and, in particular,
we can choose $a_{22}=0$.

As it can be seen from Eq. (\ref{potential}), the potential
$V$ vanishes when $z \to z_{h}$ so that
Eq. (\ref{Schrod1}) takes the form of a classical harmonic-oscillator
equation with frequency $\omega$. The solutions of such equation
are proportional to $\exp(\pm i\omega r_{\ast})$, being denoted
by $\psi_{-}$ and $\psi_{+}$. The minus(plus)-sign solution can be
interpreted as an ingoing (outgoing) plane wave at horizon $z=z_{h}$.
The wave functions $\psi_{\pm}$ can also be expressed
in terms of the normalizable and non-normalizable solutions:
\begin{equation}
\psi_{\pm}={\cal A}_{(\pm)}\psi_{2}+{\cal B}_{(\pm)}
\psi_{1}\, ,\label{ingoingoutgoingexp}
\end{equation}
where ${\cal A}_{(\pm)}$ and ${\cal B}_{(\pm)}$ are
connection coefficients associated to Eq. (\ref{Schrod1}),
and determined as functions of $\omega$ and $q$ by imposing boundary
conditions on the field solution.

\section{The glueball spectral function}

\subsection{General expression}

The Minkowskian prescription of Ref. \cite{Son:2002sd} can be used
to compute the glueball retarded Green's function $G^{R}(\omega,q)$ and,
consequently, the corresponding spectral function $\mathcal{R}=
-2 \,{\rm Im }\, G^R(\omega, q)$. First we write down the on-shell version of
the scalar field action (\ref{action0}), using the equation
of motion (\ref{eqwave1}): 
\begin{equation}
S_{\mbox{\tiny{on shell}}}=\frac{\pi^{3} L^{5}}{4\kappa_{10}^2}\int d^{4}x\sqrt{-g}
e^{-\Phi}g^{zz}\phi\partial_{z}\phi{\vert}_{_{z=0}},
\label{action1}
\end{equation}
So, we use the ``bulk to boundary propagator'' $\phi_{k}(z)$ to
decompose the scalar field $\phi(z,k)$ as
$\phi(z,k)=\phi_k(z)\phi_0(k)$. In terms of $\phi_{k}(z)$,
the action (\ref{action1}) takes the form
\begin{equation}
S_{\mbox{\tiny{on shell}}}=-\int\frac{d^{4}k}{(2\pi)^4}\phi_0(-k){\cal F}(k,z)\phi_0(k) {\vert}_{_{z=0}} .
\label{action2}
\end{equation}
According to the Minkowskian prescription \cite{Son:2002sd},
the retarded Green's function is then given by
\begin{eqnarray}
G^R(k) &\equiv& - 2 \,{\cal F}(k,z)\,{\vert}_{_{z=0}}\\
&=&\frac{\pi^3 L^5}{2\kappa_{10}^2}  \lim_{z \to 0} \sqrt{-g}
g^{zz} e^{-\Phi} \phi^{\ast}_{k}(z) \partial_z \phi_k(z)\nonumber,
\end{eqnarray}
where the propagator $\phi_{k}(z)$ satisfies an incoming-wave condition at
horizon and the normalization condition $\lim_{z\to0}\phi_{k}(z)=1$
at the boundary.

\begin{figure*}[htb]
\centering
\includegraphics[width=5.45cm,angle=270]{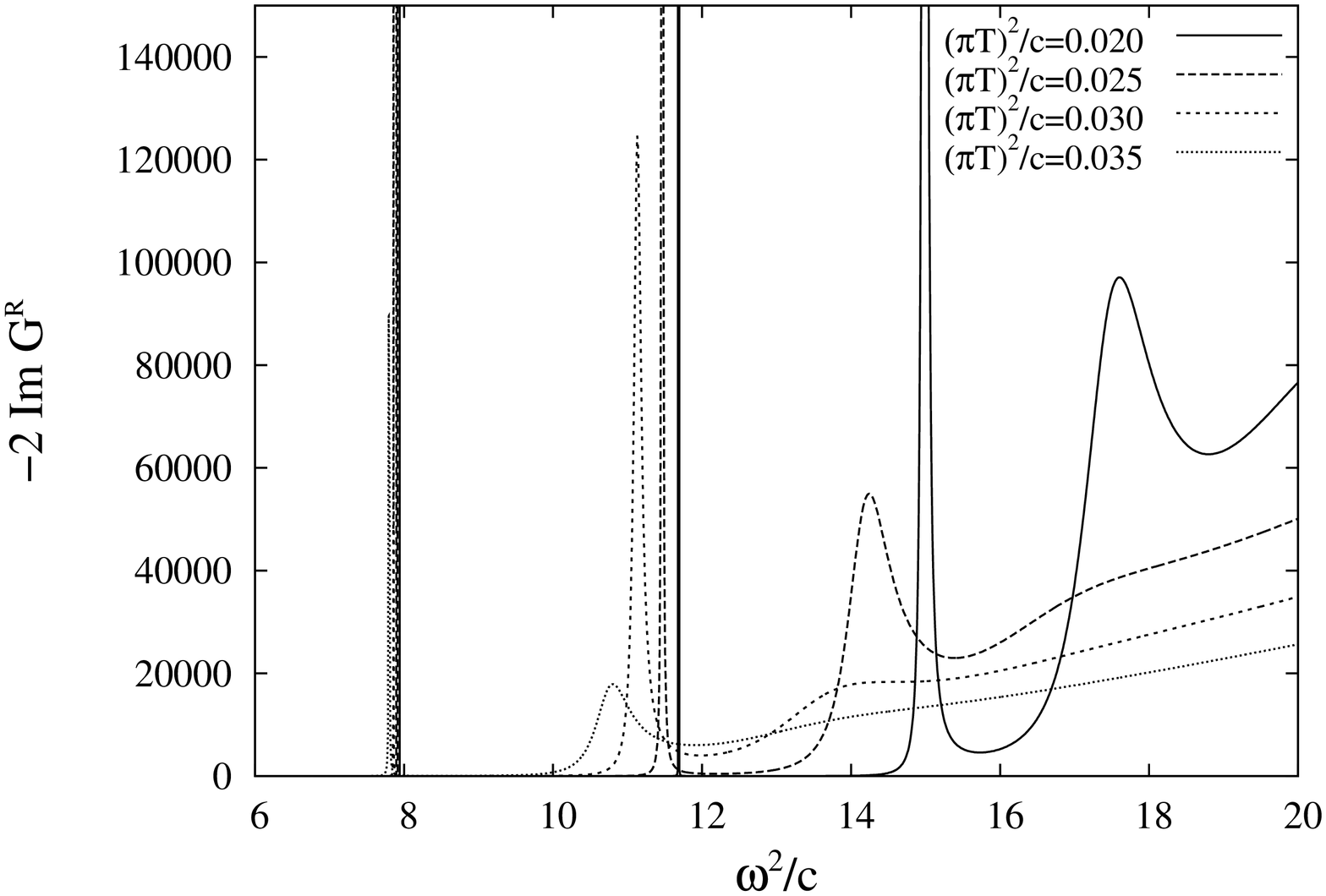}
\includegraphics[width=5.45cm,angle=270]{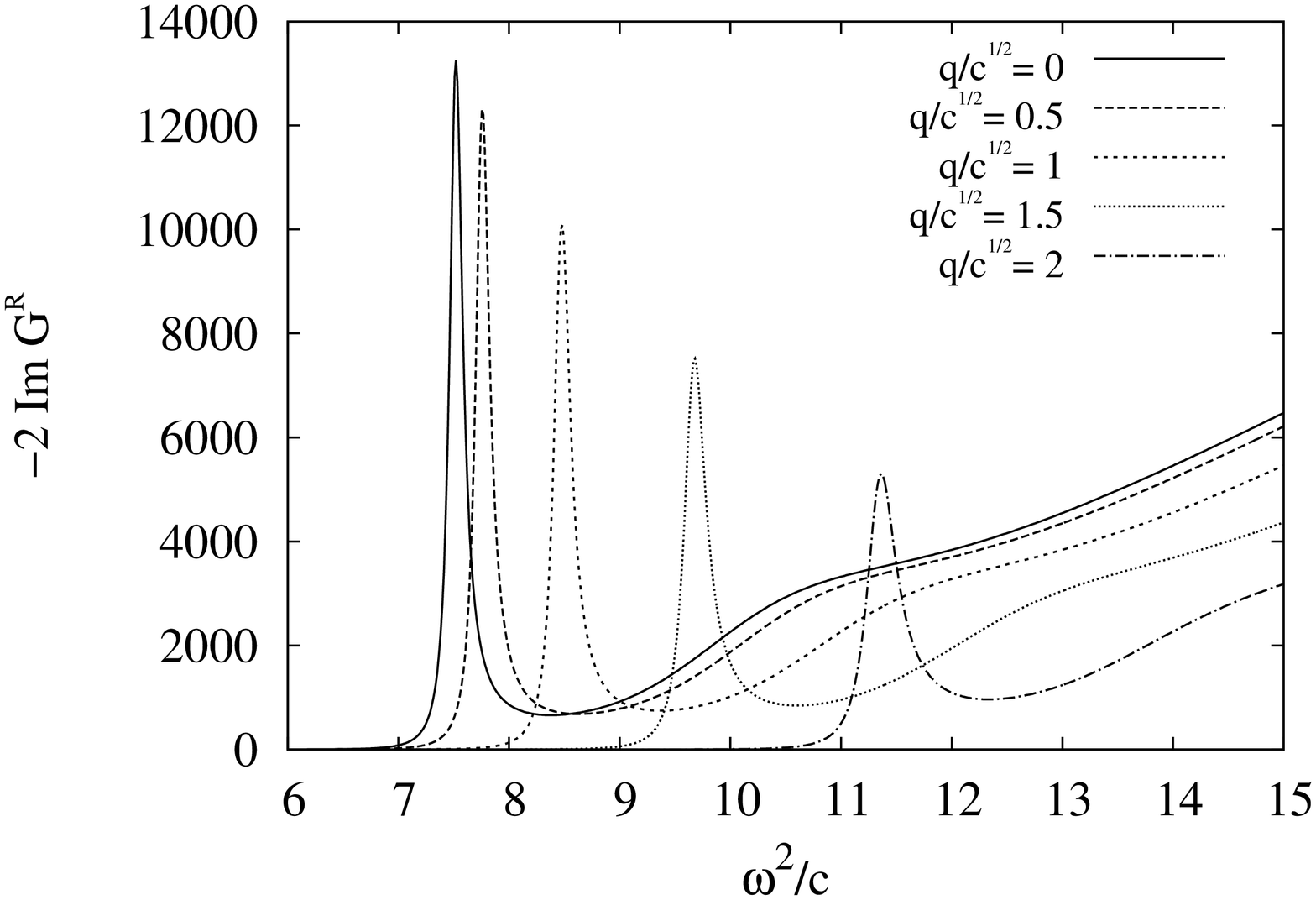}
\vspace{-0.5cm}
\caption{Left: Spectral function for $q=0$ and selected values of temperature.
Right: Spectral function for $\widetilde{T}^2=0.05$ and selected values
of the momentum. The spectral functions are in units of $N_{c}^{2}/4\pi^{2}$.}
\label{spectralfunction}
\end{figure*}

The function $\phi_{k}(z)$ can be expressed in terms of
the wave functions defined in the last section as
\begin{eqnarray}
\phi_k(z)&=&z^{3/2}e^{\frac{cz^2}{2}}\frac{\psi_{-}}{{\cal A}_{(-)}}
\nonumber\\
&=&z^{3/2}e^{\frac{cz^2}{2}}\left[\psi_{2}+
\frac{{\cal B}_{(-)}}{{\cal A}_{(-)}}\psi_{1}\right].
\end{eqnarray}


Then, using the relation $\pi^{3} L^{8}/2 \kappa_{10}^{2}=N_c^{2}/8 \pi^{2}$
and the expansions (\ref{normalizable}) and (\ref{nonnormalizable}) for
$\psi_1$ and $\psi_2$, we obtain
\begin{eqnarray}
G^{R}(k)&=&\frac{N_{c}^{2}}{8\pi^2}\left[(c+2a_{21})\epsilon^{-2}+
4b\,\mbox{ln}(c\epsilon^2)\right.\nonumber\\
&&\left.+2\left(b+a_{21}c
+a_{21}^{2}\right)\right.\nonumber\\
&&\left.+4\,{\rm Re}{\frac{{\cal B}_{(-)}}{{\cal A}_{(-)}}}
-i\,{\rm Im}{\frac{{\cal B}_{(-)}}{{\cal A}_{(-)}}}\right],
\label{greenfunction}
\end{eqnarray}
where $\epsilon$ is an ultraviolet regulator. In the above
equation, the $\epsilon$-dependent terms are ultraviolet
divergent terms that can be removed using a holographic
renormalization procedure \cite{Bianchi:2001kw}.

\subsection{Numerical results}

To find the glueball spectral function we need to compute
the imaginary part of the retarded Green's function
$G^{R}(\omega,q)$. From equation (\ref{greenfunction}) it follows that
\begin{equation}
{\cal R}(\omega,q)=\frac{N_c^2}{4 \pi^2}{\rm Im}
\frac{{\cal B}_{(-)}}{{\cal A}_{(-)}}. 
\end{equation}
The ratio ${\cal B}_{(-)}/{\cal A}_{(-)}$ can
be computed by a direct numerical integration of
equation (\ref{Schrod1}) following the steps described,
for example, in Refs. \cite{Miranda:2009uw,Teaney:2006nc}.

\begin{figure*}[htb]
\centering
\includegraphics[width=5.45cm,angle=270]{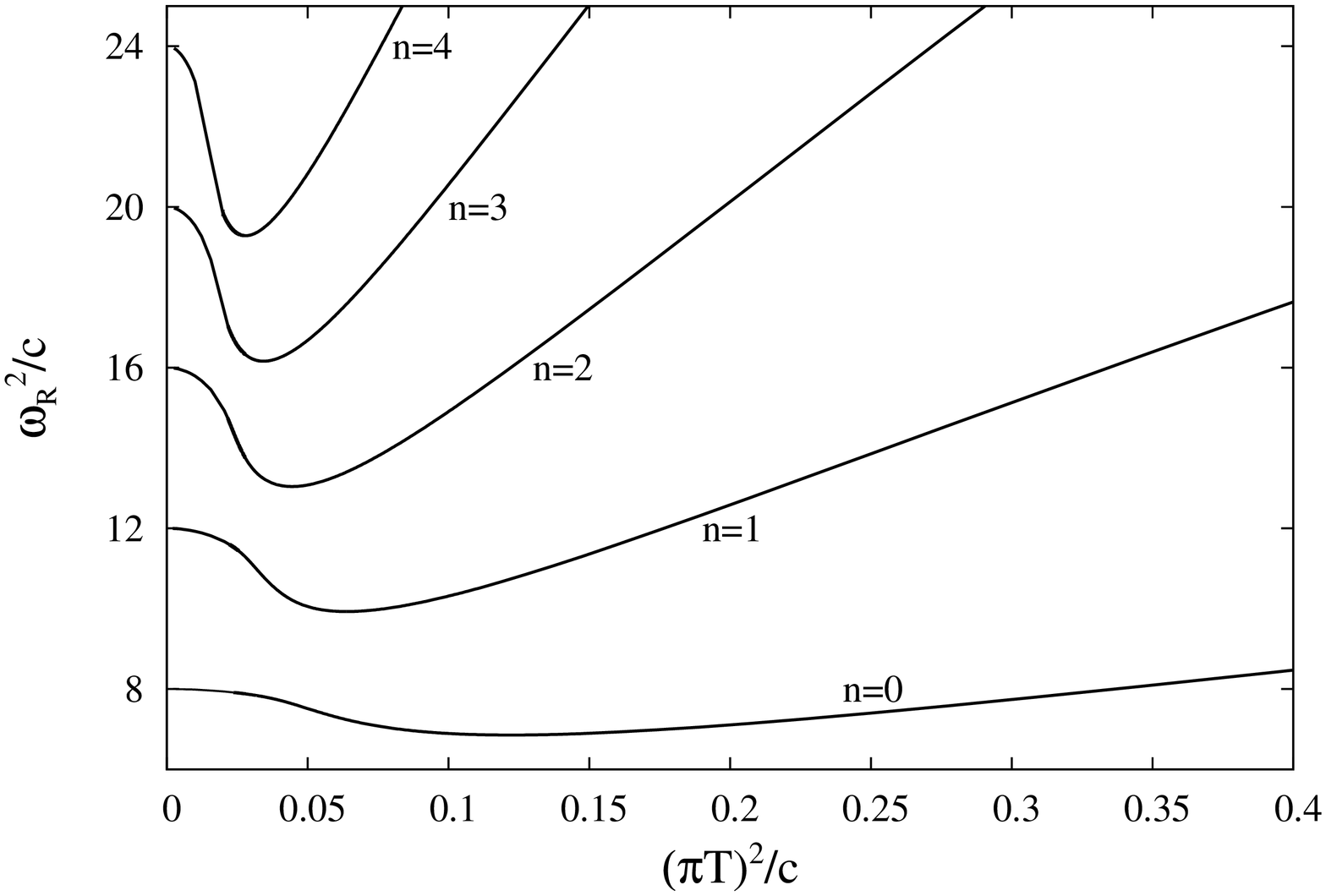}
\includegraphics[width=5.45cm,angle=270]{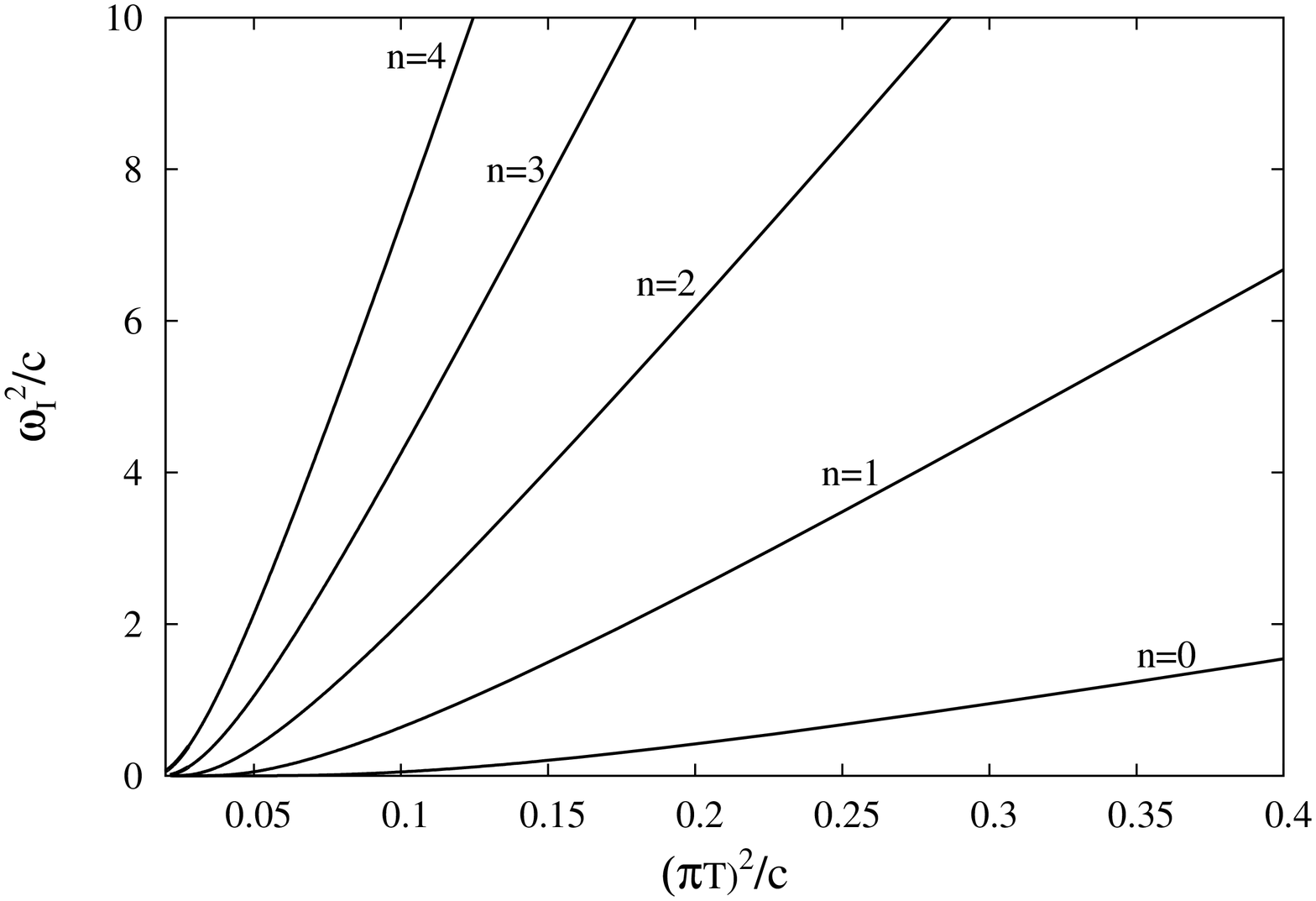}
\vspace{-0.5cm}
\caption{Numerical results for the square of the real and imaginary
parts of the QN frequencies, ${\omega}_{R}^{2}/c$ and
${\omega}_{I}^{2}/c$, for the first five quasinormal modes
$n=0,1,...,4$, with $q=0$.}
\label{quasinormal}
\end{figure*}

In figure \ref{spectralfunction}, we show the
dependence of the spectral function with the frequency
$\widetilde{\omega}=\omega/\sqrt{c}$ for some values of
temperature $\widetilde{T}=\pi T/\sqrt{c}$ and
momentum $\widetilde{q}=q/\sqrt{c}$. The various peaks
of $\mathcal{R}(\omega,q)$ correspond to the poles
of the retarded Green's function $G^{R}$ and are
interpreted as glueball states in the gauge
theory. According to the gauge/gravity duality,
these poles are associated with the frequencies of
the black-hole quasinormal modes.

As it can be seen in the left panel of figure \ref{spectralfunction},
the number of glueball states and their lifetimes (the inverse of
the half-width) decrease with the temperature. In fact, there is a
temperature value ($\widetilde{T}^{2}\sim 0.1$) from which
on there are no more peaks in the spectral function,
characterizing a {\textit{glueball melting}}.
For a fixed value of temperature $\widetilde{T}$, the position and the width
of the peaks increase with the momentum $\widetilde{q}$, as shown in
the right panel of figure \ref{spectralfunction}.

\section{From quasinormal modes to glueballs}

Two different methods have been used to compute the scalar-field
QNM frequencies of the AdS black-hole spacetime (\ref{background}).
The first one is a series expansion method \cite{Horowitz:1999jd},
which reduces the problem to finding roots of a polynomial,
and it is suitable to find QNM frequencies in the high and intermediate
temperature regimes. The second method is based on the computation
of Breit-Wigner ressonances \cite{Berti:2009wx}, and it is efficient to
investigate the spectrum for very low temperatures. The approaches
are complementary and present a very good concordance in a wide
range of parameter space \cite{Miranda:2009uw}.

We show in Fig. \ref{quasinormal} the real and imaginary parts
of the QNM frequencies as function of the temperature for the first
five modes $n=0,\,1,\,...,\,4$ with $q=0$. In the limit $\widetilde{T}\rightarrow 0$,
the quasinormal modes tend to the spectrum of glueballs at zero temperature: 
$m_{{G}_{n}}^{2}=4c(n+2)$.
As the temperature increases the imaginary part of the
frequency $\widetilde{\omega}_{I}$
also increases, while the real part $\widetilde{\omega}_{R}$ decreases
with $\widetilde{T}$. It means that the mass and the decaying time
of the glueball excitations decrease with $\widetilde{T}$ for very
low temperature values. In the intermediate- and high-temperature
regimes, the effect of the dilaton background field is small,
and the QNMs are essentially the same as those of the black hole
solution with $\Phi=0$ \cite{Starinets:2002br,Nunez:2003eq,Kovtun:2005ev,Morgan:2009pn}. 
In particular, after reaching some minimum value the real part $\widetilde{\omega}_{R}$ 
increases with temperature tending asymptotically to a linear dependence. 
In fact, the fractional differences between the cases with and without
dilaton reach a value of the order of $0.15$ for
$\widetilde{T}=1$, and tend asymptotically to zero
in the limit of high temperatures  \cite{Miranda:2009uw}.

\section{Final comments}

In this article we review recent progress \cite{Miranda:2009uw}
on the study of the spectrum of scalar glueballs at
finite temperature. The AdS/QCD soft-wall model and
a black-hole background spacetime have been used to
investigate thermal effects on glueball excitations
of the plasma phase. As it was shown in Ref. \cite{Miranda:2009uw}
the black-hole space are associated to an unstable,
metastable or stable phase of the dual theory
according with the value of the temperature parameter
$\widetilde{T}=T/\sqrt{c}$.
It was noticed the existence of peaks in the spectral function,
corresponding to glueball states,
only for temperatures lower than $\widetilde{T}^2=0.1$, a regime for
which the black hole is in an unstable phase. So according
to our results, we do not expect to find glueballs in the
hot plasma phase ($\widetilde{T}^2>2.39$) of the QCD.
Similar results were recently obtained for scalar \cite{Colangelo:2009ra}
and vector \cite{Fujita:2009wc} mesons in the soft-wall model.
It would be interesting to extend these computations to
other fields and other AdS/QCD models, like
those of Refs. \cite{Gursoy:2007cb,Gursoy:2007er,dePaula:2008fp}.
Such analysis could reveal what aspects of our results are specific
of the holographic model considered here and what aspects
represent general characteristics of glueball excitations
in the dual plasma.


\section*{Acknowledgements}

ASM and NRFB are grateful to the organizers of the
``Light-Cone 2009: Relativistic Hadronic and Particle Physics''
and specially to Tobias Frederico
for their hospitality. The authors are partially supported by CNPq,
Capes and Faperj, Brazilian agencies.


\end{document}